# EMERGENCY RESPONSE COMMUNICATIONS AND ASSOCIATED SECURITY CHALLENGES

Muhammad Ibrahim Channa[1] and Kazi M. Ahmed[2]

[1]Information and Communication Technologies, Asian Institute of Technology, Thailand
muhammad.ibrahim.channa@ait.ac.th
[2]Telecommunications, Asian Institute of Technology, Thailand
kahmed@ait.ac.th

## ABSTRACT

*The natural or man-made disaster demands an efficient communication and coordination among first responders to save life and other community resources. Normally, the traditional communication infrastructures such as landline or cellular networks are damaged and don't provide adequate communication services to first responders for exchanging emergency related information. Wireless ad hoc networks such as mobile ad hoc networks, wireless sensor networks and wireless mesh networks are the promising alternatives in such type of situations. The security requirements for emergency response communications include privacy, data integrity, authentication, key management, access control and availability. Various ad hoc communication frameworks have been proposed for emergency response situations. The majority of the proposed frameworks don't provide adequate security services for reliable and secure information exchange. This paper presents a survey of the proposed emergency response communication frameworks and the potential security services required by them to provide reliable and secure information exchange during emergency situations.*

## KEYWORDS

*Emergency response communication, Mobile ad hoc networks, Wireless sensor networks, Wireless mesh networks, Security, Reliability.*

## 1. INTRODUCTION

It is a great challenge for modern societies to cope with crisis situations that arise due to natural or man-made disasters. Emergency or crisis management refers to various activities such as immediate response, recovery efforts, disaster mitigation, and preparedness efforts for reducing the impact of possible future disasters. The timely access to desired information by intended person or rescue organization is vital for successful emergency management operations. Depending upon the intensity and coverage area of a disaster, it might be a multi-organizational operation involving government authorities, public authorities, volunteer organizations and the media. These entities work together as a virtual team to save lives and other community resources [1]. In emergency situations, the availability of telecommunication services is of great importance. These systems provide a means for communication among first responders and affected people. During Hurricane Katrina, several wireless base stations were taken out and various communication cables were damaged. The remaining parts of the network were not able to provide adequate communication services to first responders [2]. Mobile ad hoc networks [3], wireless sensor networks [4] and wireless mesh networks [5] are commonly used as communication means in such type of situations. These systems are easily deployed without need of any existing telecommunication infrastructure. These networks automatically configure when new nodes join or leave the network dynamically.

Reliable and robust communication is vital for successful emergency management operations [2]. The emergency response operation is hierarchical in nature, where information and instructions flow along a chain of entities. For example, disaster site videos and other related information is transferred from first responders to their local command centres and respective remote headquarters. Similarly, the instructions and site maps are communicated from remote headquarters to local command centres and first responders for efficient emergency management. Efficient emergency management requires the continuous flow of bi-directional information among first responders and emergency management headquarters. If there is some interruption in this flow, it may cause mismanagement of emergency response efforts.

During emergency situations, the deployed ad hoc communication network might be vulnerable to faults and security threats [1]. In large scale emergency management operations, the participating nodes may belong to several rescue teams and there is a possibility that a subset of nodes may experience faults or demonstrate selfish/malicious behaviour. The faults might occur due to some damage during field level emergency management operations. The emergency response network deployed during a man-made disaster such as 9/11 may be more vulnerable to malicious threats for creating more panic situation. The selfish nodes may drop the routing and data messages for saving their battery life. The malicious nodes may try to temper with the routing and data messages or inject erroneous messages in the network. These behaviours affect the reliability of emergency response networks and might result into mismanagement of emergency response efforts.

The rest of the paper is organized as follows. Section 2 describes the most common security challenges for emergency response communications. An overview and potential security challenges for ad hoc, sensor and mesh networks for emergency situations are described in section 3, section 4 and section 5 respectively. Section 6 presents comparative analysis of the proposed schemes based on the security services offered by them and section 7 concludes the paper.

## 2. SECURITY CHALLENGES FOR EMERGENCY RESPONSE COMMUNICATIONS

Possible security challenges for emergency response communications include privacy or confidentiality, data integrity, authentication, key management, and access control [2]. The following subsections describe these challenges in brief.

### 2.1. Privacy

The privacy or confidentiality ensures that the information is available only to the authorized persons and is achieved by applying cryptographic techniques. In emergency response communications, various rescue teams use the same network and it is required that each organization receives the information intended for it. The data should be encrypted in order to provide right information to the right person [17]. For example; the patient's medical history should be available in comprehensible format to the medical teams only. The emergency communication networks use resource constrained mobile devices and require a light weight cryptographic system [12]. The traditional public/private key cryptography is not suitable for such networks as it requires more computations and consumes more battery power.

### 2.2. Data Integrity

Data integrity confirms the completeness or accuracy of the transmitted information and ensures that no one has changed it on the way. The deliberate or accidental modification of the sensitive information, such as patient's medical history, accessed at the emergency site from some remote hospital database might create more panic situation. The alteration of the information representing instructions from command headquarters to first responders might also result into mismanagement of emergency response efforts.

### 2.3. Authentication

Authentication is the process of verifying the identity of someone or something. An authentication mechanism should ensure that only the right persons access the information depending upon the application requirements [17]. When the emergency network is used to transfer medical history, the user authentication is required for providing information to the legitimate person [9]. A flexible security model is required for authenticating medical staff when accessing patient history from hospital database. The medical teams should also be able to handoff access rights to other peers when required [12].

### 2.4. Key Management

It is worthwhile to exchange security credentials (such as, keys, digital certificates) efficiently during emergency response communications. These security credentials help first responders in performing certain security functions such as authentication and encryption. In a large scale emergency response scenario, it is not possible to assume that all the rescue organizations have exchanged security credentials before joining the emergency response network. There is a need of developing an efficient key distribution and management scheme for such type of situations [18].

### 2.5. Access Control

In life threatening situations, such as emergency response, the legitimate user must have immediate access to the required information that might save someone's life [18]. Access-control policies provide access to critical data, such as patient's healthcare information, under various constraints and environments. Normally, the access is provided depending upon the identity, time and location constraints. For example; a physician can access patient's information between 9.00 AM to 5.00 PM within the boundaries of the hospital. This prevents a physician to access the same information from emergency site. It is required to develop an adaptive access-control system suitable for emergency situations. The information should also be available for access to the legitimate users all the times.

## 3. SECURITY CHALLENGES FOR MOBILE AD HOC NETWORKS IN EMERGENCY RESPONSE

An emergency and disaster response model is presented, which makes use of the ambient intelligence (AmI) technologies to support communications among participating rescue teams such as police, fire fighters and ambulance services [6]. The ambient intelligence technologies provide adaptive and assistive services to users by assuming a great number of interoperating devices such as sensors, actuators and other devices performing storage, processing and communication of data. Figure 1 describes the scenario. The hospitals, police cars, ambulances, fire fighters and medical teams are integrated in to a single virtual team performing disaster management operations. The system uses body area network (BAN), personal area network (PAN), mesh network, ad hoc network, sensor network, cellular network, terrestrial trunked radio (TETRA) network and global network as communication means. The proposed system is a conceptual scenario for future emergency response communications. As shown in figure 1, the information may flow through heterogeneous networks under different administrative controls. It is possible that some nodes might intercept secret information, such as patient's history, or generate fake information. Privacy and authentication are the key requirements in this type of scenario for reliable communications. The access to patient's history at some remote hospital from the emergency site also requires an adaptive access control mechanism. The data integrity is also crucial as the emergency related information passing through heterogeneous networks may also be modified intentionally or accidentally.

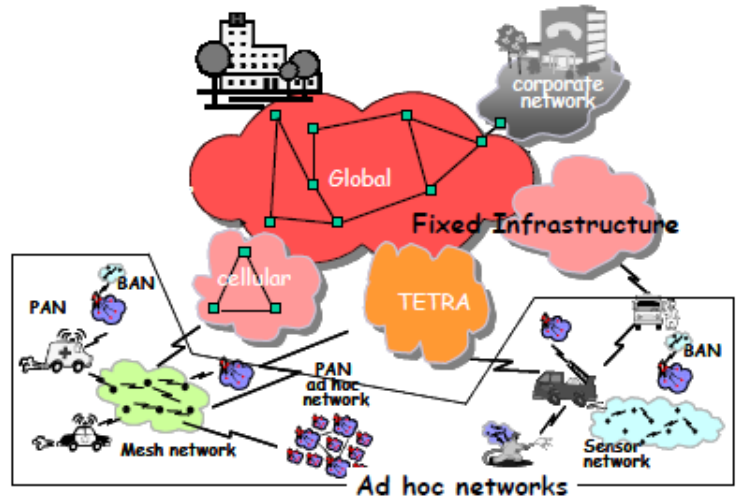

Figure 1: Ambient intelligence technologies for emergency support [6]

A project had been initiated in Romania to develop a disaster response communication system for providing services such as quick information and alert dissemination to people, collaboration and coordination among rescue teams, information aggregation and personal assistance in emergency situations [7]. The system uses modern mobile communication technologies to provide a useful instrument to the government for controlling emergency situations. The general architecture of the system is presented in figure 2. This is a multi-step model comprising of four steps or modules. In first step, the disaster warnings are communicated to potentially affected people through traditional communications infrastructure, such as cellular phones. This creates awareness among people so that they may make appropriate arrangements to reduce the potential loss of life and other resources. In second step, the system communicates the aggregate information from rescue teams to decision makers for efficient and timely decision making. In third step, rescue teams collaborate and coordinate with each other and in the final step; affected people receive assistance in the shape of building plans, directions and possible exit routes in damaged buildings. It is apparent from figure 2 that various remote entities communicate with each other and exchange emergency related information. Authentication is required during the propagation of warnings to potentially affected people for avoiding rumours. An adaptive access control mechanism is also necessary to have immediate access to some life saving information at the emergency site. The privacy and data integrity are also essential for providing secrecy and accuracy to the sensitive information such as medical history of affected people.

An emergency communication system, called digital ubiquitous mobile broadband OLSR network (DUMBONET) is proposed for exchanging real time multimedia information in search and rescue operations in disaster affected areas [8]. It connects various disaster affected areas and command centre through long delay satellite links. Every site has its own mobile ad hoc network for communications among first responders. The mobile node in DUMBONET may be a light weight laptop computer or a personal digital assistant (PDA). The system allows every rescuer and command headquarter to communicate and coordinate by using video, voice and short messages. The DUMBONET uses geostationary satellite providing broadband connectivity among disaster sites and central command headquarter and is shown in figure 3.

The key security challenges for DUMBONET include privacy, authentication, data integrity and access control. The privacy is required for secure exchange of secret information through multiple emergency sites. Authentication is required for reliable communications among head command centre and various disaster areas. The flow of information through multiple disaster areas requires the implementation of a data integrity mechanism that identifies the deliberate or accidental information modifications. An adaptive access control mechanism is also vital for quick access to life saving information at the emergency site.

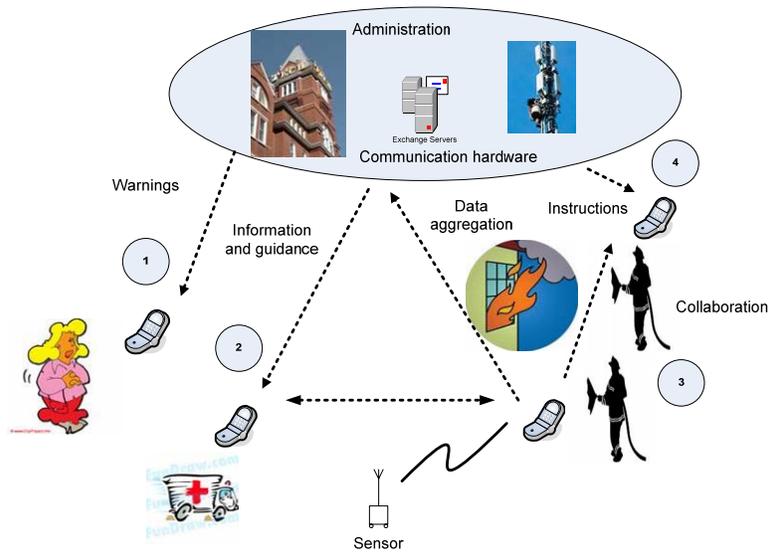

Figure 2: Government controlled citizen's safety system [7]

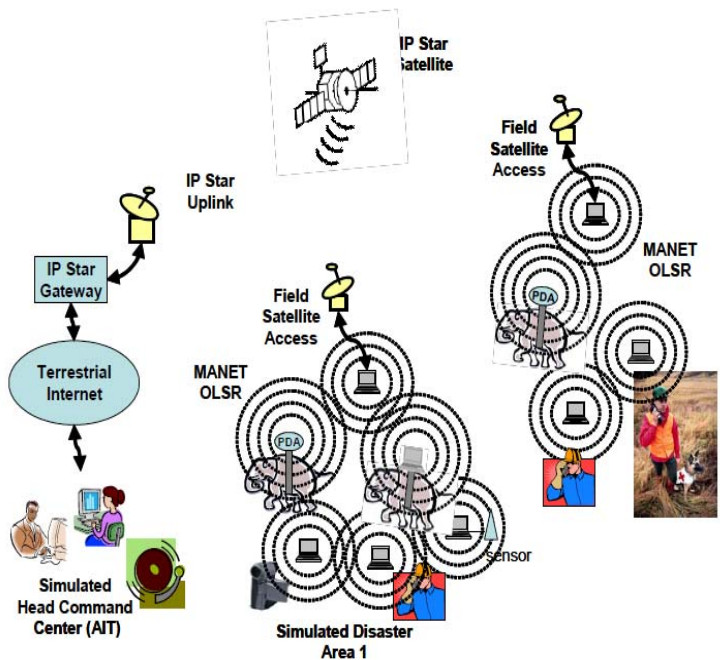

Figure 3: DUMBONET test bed [8]

A disaster response communication framework, called self-organizing ad hoc network for first responders (SAFIRE) is proposed to facilitate information exchange among first responders [9]. It provides a decentralized cognitive radio based approach for supporting interoperability, better performance and usability. It provides publish-subscribe service, where some nodes publish the information and others subscribe to access the same. It comprises of four modules namely publish-subscribe module, routing/forwarding engine, radio module and policy module. The publish-subscribe module is used for storing and accessing information contents. The routing/forwarding engine is responsible for finding routes and forwarding data over them. The radio module manages the cognitive radio configuration and the policy module forms rules and regulations for all integrated modules. The SAFIRE system is shown in figure 4. The SAFIRE is a content-based emergency response system, where nodes publish and subscribe emergency related information for efficient emergency management operations. The key security challenges for SAFIRE include authentication, data integrity and access control. Authentication is required to verify the identity of the source publishing the information to avoid incorrect information retrieval. The information might be modified on the way, so a data integrity mechanism is also required. The medical teams at emergency sites may require an adaptive access control mechanism to quickly access the life saving information from some remote hospital database. The privacy is also essential for secure exchange of sensitive data such as patient's history.

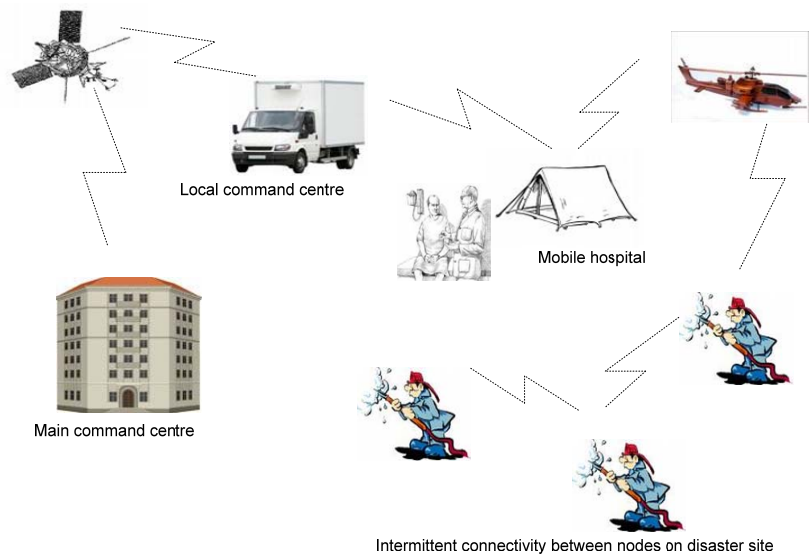

Figure 4: SAFIRE system [9]

A public safety and disaster recovery communications system is proposed to facilitate communications among first responders during emergency situations [2]. The framework comprises of four types of communication networks including personal area network (PAN), incident area network (IAN), jurisdiction area network (JAN) and extended area network (EAN) as shown in figure 5. A PAN interconnects various devices carried by individual rescue team members. An IAN is a network that is established temporarily at the incident site among first responders. A JAN is a permanent network typically installed by public safety agencies or municipalities to provide wide area connectivity. An EAN provides wide area connectivity among various regional, state and national public safety networks. The proposed public safety system is a conceptual framework for supporting information exchange among first responders and emergency headquarters during emergency management operations. The privacy is required to hide the secret information travelling through heterogeneous networks from unauthorized users. Data integrity is essential because the information travelling through heterogeneous networks managed by different entities might be changed on the way. Authentication is required for identifying the legitimate users for authentic information exchange at all network levels.

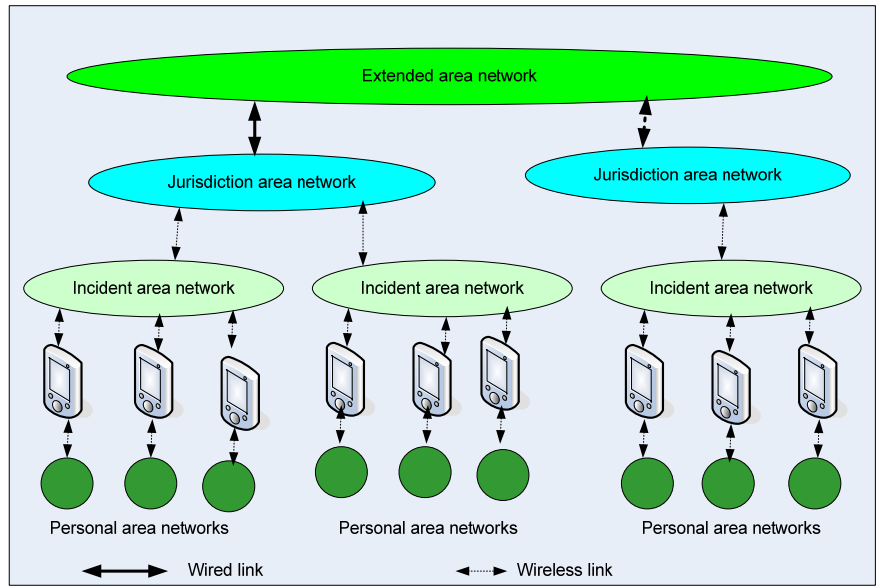

Figure 5: Public safety communications system [2]

A post-disaster telecommunication network using cognitive agents has been proposed in [10]. The proposed framework provides immediate post-disaster communications facility to the surviving mobile nodes and enables them to communicate among them and disaster information centre (DIC). The communication among them takes place through some surviving base stations and satellite links. Figure 6 describes the proposed framework. Four cognitive agents, namely, emergency identifier (EI) agent, topology explorer (TE) agent, information gatherer (IG) agent and information disseminator (ID) agent are available in every mobile node. The EI agent roams around the environment and collects information about the current network state and identifies emergency situations. The TE agent monitors the dynamic network topology when nodes join and leave the network. The IG agent gathers information from the network and other agents having different parent nodes. The ID agent is initiated by a mobile node which has some information to disseminate about the environment and surviving neighbouring nodes. The key security challenge for the proposed framework is authentication. Without proper authentication, a compromised node might generate false emergency information creating panic situation. While exchanging emergency related information, every agent should verify the identity of the information source before taking appropriate action.

## 4. SECURITY CHALLENGES FOR WIRELESS SENSOR NETWORKS IN EMERGENCY RESPONSE

A distributed search and rescue system comprising of humans, sensors and robots has been proposed for emergency situations [11]. This system is especially designed for finding exit routes for affected people from damaged buildings during emergency situations. The sensors collect the information of their local environments and exchange it for further analysis. The maps are generated from this information and used by humans and robots to navigate the target while avoiding dangerous areas. The proposed system is shown in figure 7. The proposed system was experimented by using mobile robots equipped with sensors for detecting fire gradients. When the robots reach near fire, they stop there and move towards the building exit after receiving appropriate instructions. The proposed system generates the exit route maps depending upon the information provided by the sensors. It is vital to validate the information in order to generate the correct maps. Authentication and data integrity are crucial for the proposed scheme. Authentication is required for validating the information source and data integrity is necessary to validate the completeness and accuracy of the received information, which is necessary for generating correct exit route maps.

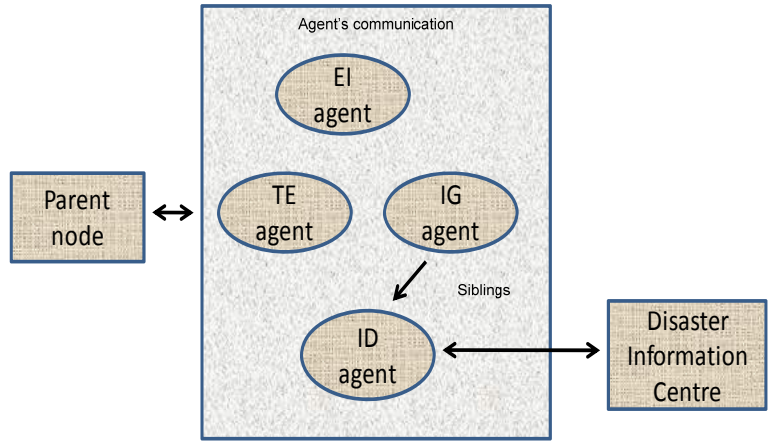

Figure 6: Post-disaster agent based communications [10]

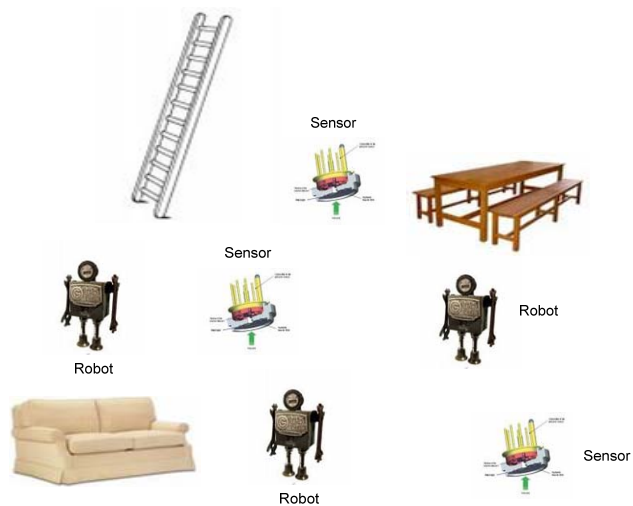

Figure 7: Sensors and robots in search operation [11]

A sensor-based ad hoc network called CodeBlue is proposed for emergency medical care consisting of vital sign sensors, personal digital assistants (PDAs) and personal computer (PC) systems [12]. The system enables medical teams to assess the condition of patients at the emergency site and ensures seamless transfer of data among medical personnel. This architecture provides reliable data transfer and a decentralized security system as shown in figure 8. It provides handoff of credentials across rescue personnel and seamless access control across patient's transfer. The CodeBlue uses publish/subscribe model for data delivery. The sensor nodes publish vital signs, location information and identities. The physicians subscribe to this information through PDAs and PC systems to monitor patient's status. The major security services offered by CodeBlue include authentication and encryption. It is desired to develop light-weight cryptographic techniques for resource constrained mobile nodes in CodeBlue. The encryption service also provides confidentiality of critical information, such as vital signs of patients.

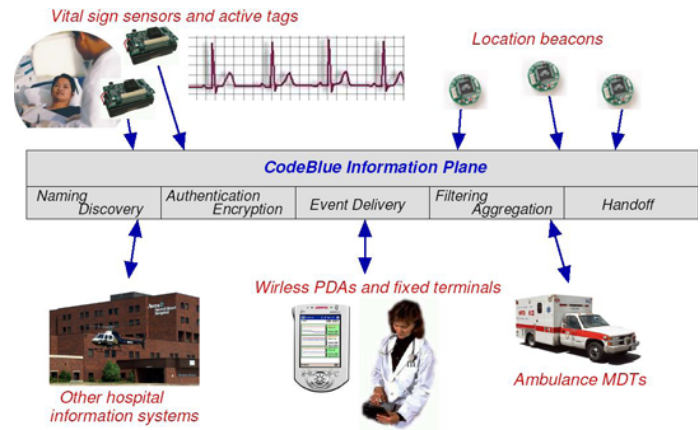

Figure 8: CodeBlue architecture [12]

An efficient disaster management communication and information system (DMCIS) is proposed to generate and communicate pre-disaster warnings to potentially affected people [13]. The proposed architecture comprises of wireless sensor networks and mobile ad hoc networks for collecting data about disaster hotspots and emergency situations. The framework is divided into four levels, where each level interacts with other levels in providing related disaster information at the appropriate place. At the first level, the sensor nodes deployed at sensor and data collection centre (SDCC) monitor the vital parameter changes (i.e., level of water in the river for flood warning) and then send them to the SDCC. The data acquisition process at SDCC is shown in figure 9. At level 2, an ad hoc network is formed for data transmission from SDCC to data processing center (DPC). This communication is facilitated through a vehicle mounted mobile access point (MAP) using Wi-Fi technology. At level 3, the acquired data at the DPC is processed. A DPC also maintains detailed past records related with disasters for the areas associated with it. Once the DPC processes the data, it is transferred to the central data centre (CDC) through wired or wireless communications. At level 4, the CDC checks the similarities of the information with past records of previous disasters. When CDC finds high probability of a disaster, it requests decision and command centre (DCC) to take responsive actions. The DCC in turn calls the police, fire, medical and other emergency services to reach at the disaster site. To warn people about the potential disaster, the DCC sends information to the local mobile phone service providers to disseminate the warnings via short message services (SMS) to their subscribers. The main security challenges for DMCIS include authentication, and data integrity. Authentication is required to validate the source of the information as the sensor nodes at SDCC are installed openly and might be compromised. The data integrity is required to validate the completeness and accuracy of the information sent by the sensors for correct prediction of the potential emergency situation. The authentication and data integrity services should be available at all levels of the proposed scheme.

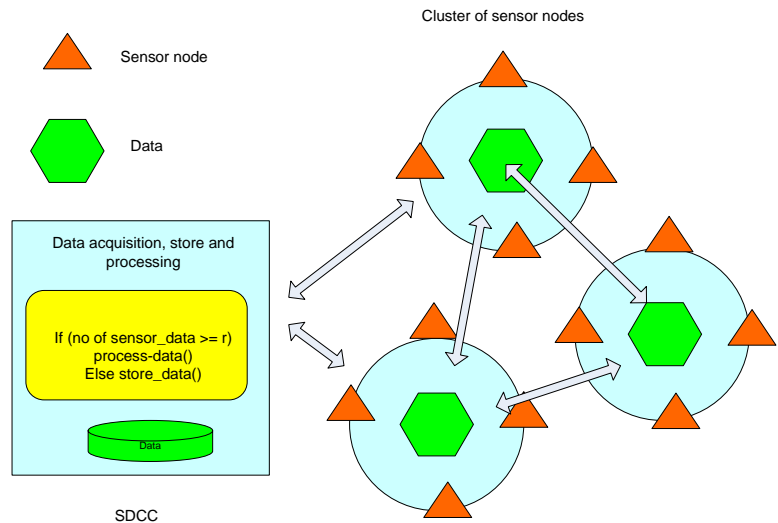

Figure 9: Data collection at SDCC [13]

## 5. SECURITY CHALLENGES FOR WIRELESS MESH NETWORKS IN EMERGENCY RESPONSE

A crisis management system, called GeoBIPS, has been proposed to provide timely and up-to-date information during crisis situations [14]. The GeoBIPS is a self-forming broadband wireless mesh network used by reconnaissance team members (RT) and commanding officer (CO). The relay network is serving as an interface between RT and CO passing information in both directions. This enables CO to acquire up to date information from RT members, convey information to the crisis centre for decision making and send useful data and instructions to RT members. The GeoBIPS architecture is shown in figure 10. The portable access router (PAR) receives digital video data and sends it over a wireless interface to CO and embedded server. The mobile access router (MAR) connects to the crisis centre by using a local hotspot, general packet radio service (GPRS) or a universal mobile telephone system (UMTS) connection. The security is provided through IP security (IPSec) tunnel between MAR and PAR for secure voice and video communications. The MAR is provided with pre-shared authentication key used to sign all OLSR routing messages to prevent malicious nodes from entering the network. The privacy is achieved through IPSec service.

A ballooned wireless mesh network is proposed to provide reliable communication means for exchanging information relating to disaster area, resident's safety and relief services during emergency situations [15]. The system comprises of multiple ballooned wireless nodes, which are hanged 40 meters above in the sky, fixed access point, mobile notebook PCs and wireless IP phones as shown in figure 11. The network is connected to the Internet and disaster information server (DIS) through fixed access point. It is apparent from figure 11 that the main source of communication among first responders at the emergency site and the disaster information server is the ballooned nodes. If some ballooned node is compromised, it might create communication interruptions or generate fake information to be exchanged between first responders at emergency site and disaster information server. Authentication is required to overcome this problem. Data integrity also counts as the exchange of incomplete and inaccurate information between emergency site and disaster information server might result into mismanagement of emergency response efforts. Privacy is also essential for the exchange of secret information among first responders and emergency information centre.

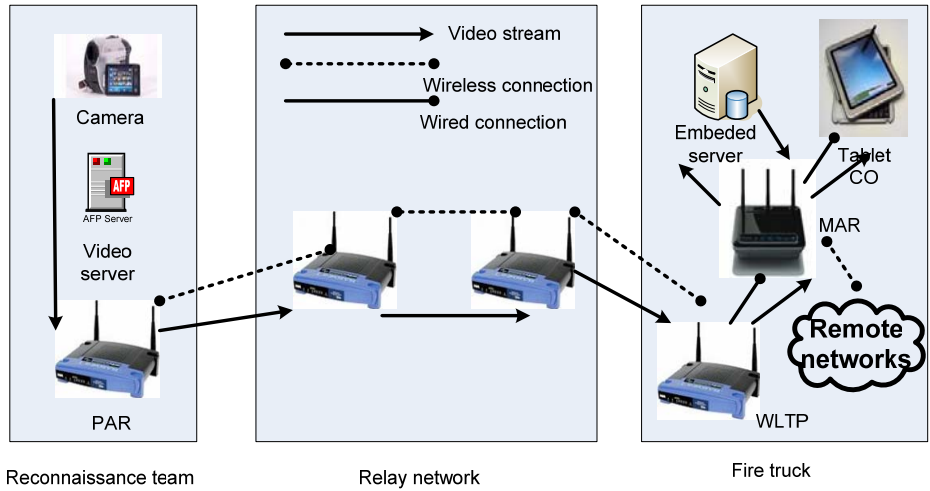

Figure 10: GeoBIPS architecture [14]

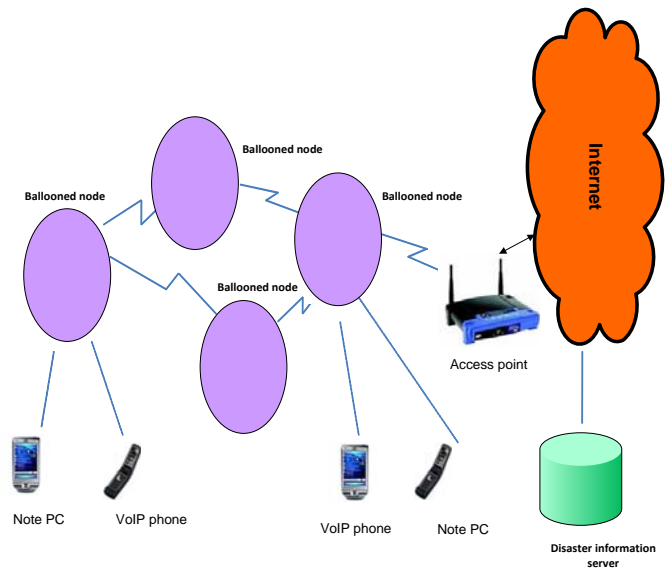

Figure 11: Ballooned emergency response network [15]

## 6. COMPARATIVE ANALYSIS OF THE PROPOSED EMERGENCY RESPONSE FRAMEWORKS

It has been observed from section 3, section 4 and section 5 that the majority of the proposed emergency response communication systems do not provide adequate security services for reliable and secure information exchange. In this section, we provide a comparative analysis of the proposed schemes based on the security services offered by them as shown in Table I. It is mentioned in Table I that the proposed schemes except CodeBlue and GeoBIPS don't provide any security service for reliable and secure information exchange. The authentication is only provided by the CodeBlue and GeoBIPS schemes, so it is possible to validate the identity of the information source. If the source is reliable, the information might be entertained, otherwise it might be ignored. For the rest of the schemes, a malicious node might impersonate a reliable node and generate fake information that might be harmful for the emergency response efforts. The privacy service is also provided by the CodeBlue and GeoBIPS only, so a malicious user or node cannot intercept the secret information on the way in those schemes. For other schemes, it is an easy task for a malicious node to intercept and misuse the secret information. The lack of an adaptive access control scheme in ambient intelligence framework, government controlled safety system, DUMBONET, and SAFIRE schemes prevents a physician to access the medical history information of the patients at the emergency site that might result into the loss of their lives. The data integrity check is also desired for all the schemes except CodeBlue and GeoBIPS for identifying the correctness and completeness of the received information, as an incomplete and inaccurate information might be dangerous. If the participating rescue teams belong to different rescue organizations, then a dynamic key distribution and management scheme is also required for authentication and encryption services.

Table I: Security services for emergency response communications

| Category | Proposed Scheme | Available Security Services | Required Security Services |
|---|---|---|---|
| Mobile Ad Hoc Network | Ambient Intelligence Framework | None | Privacy, Data Integrity, Authentication, Access Control |
| | Government Controlled Safety System | None | Authentication, Access Control, Privacy, Data Integrity |
| | DUMBONET | None | Privacy, Data Integrity, Authentication, Access Control |
| | SAFIRE | None | Authentication, Privacy, Data Integrity, Access Control |
| | Public Safety System | None | Privacy, Data Integrity, Authentication |
| | Post-Disaster Communication System | None | Authentication |
| Wireless Sensor Network | Search and Rescue System | None | Authentication, Data Integrity |
| | CodeBlue | Authentication, Privacy | More lightweight cryptographic schemes |
| | DMCIS | None | Authentication, Data Integrity |
| Wireless Mesh Network | GeoBIPS | Authentication, Privacy | None |
| | Ballooned Emergency System | None | Authentication, Data Integrity, Privacy |

## 7. CONCLUSIONS

This paper presents a survey of various communication frameworks for emergency response along with their security requirements. The majority of the proposed schemes lack the security services required for reliable and secure information exchange. The key security services include privacy, data integrity, authentication, key management and access control. In real life, there is a possibility that some nodes might demonstrate malicious behaviour by intercepting secret information or impersonate a reliable node and generate fake information. These behaviours affect the reliability of the emergency response networks and might result into mismanagement of emergency response efforts. It is required that the key security services may be incorporate in the proposed frameworks so that the reliable and secure information exchange might be made possible.

**Authors**

**Muhammad Ibrahim Channa**[1] completed B.S in Computer Science from University of Sind, Jamshoro, Sindh, Pakistan and M.S in Information Technology from National University of Sciences and Technology, Islamabad, Pakistan. Recently, he is pursuing PhD degree at Asian Institute of Technology, Thailand in Information and Communications Technologies. He is currently employed at Quaid-e-Awam university of engineering, science and technology, Nawabshah, Sindh, Pakistan as assistant professor, department of Information Technology.

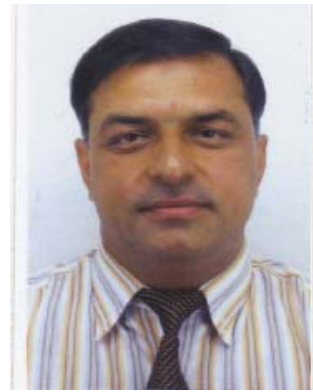

**Kazi M. Ahmed**[2] completed M.Sc. Eng., Electrical Engineering, Institute of Communications, St. Petersburg (Leningrad), Russia and Ph.D. from University of Newcastle, NSW, Australia. Currently he is working as a professor, Telecommunications, Asian Institute of Technology, Thailand. His research interests include wireless systems and networks, disaster warning and post-disaster communications, propagation and channel modelling in mobile communications, multiple access techniques and protocols, satellite communications, antenna array processing and signal processing. He is a member of IEEE and IEICE Japan.

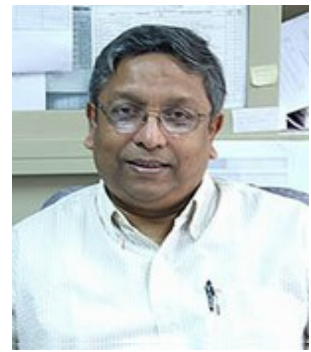